\documentclass[aps,prd,eqsecnum,preprint,tightenlines,nofootinbib,showpacs]{revtex4-1}
\usepackage{graphicx,latexsym}

\def\bea{\begin{eqnarray}}
\def\eea{\end{eqnarray}}
\def\be{\begin{equation}}
\def\ee{\end{equation}}

\newcommand{\Pminus}{{\cal P}^-}

\begin{document}

\title{Two-dimensional light-front $\phi^4$ theory \\
in a symmetric polynomial basis
}
\author{Matthias Burkardt}
\affiliation{Department of Physics \\
New Mexico State University \\
Las Cruces, New Mexico 88003}
\author{Sophia S. Chabysheva}
\author{John R. Hiller}
\affiliation{Department of Physics and Astronomy\\
University of Minnesota-Duluth \\
Duluth, Minnesota 55812}

\date{\today}

\begin{abstract}

We study the lowest-mass eigenstates of $\phi^4_{1+1}$ theory
with both odd and even numbers of constituents.  The calculation
is carried out as a diagonalization of the light-front Hamiltonian
in a Fock-space representation.  In each Fock sector a fully
symmetric polynomial basis is used to represent the Fock wave function.
Convergence is investigated with respect to the number of basis 
polynomials in each sector and with respect to the number of sectors.
The dependence of the spectrum on the coupling strength is used
to estimate the critical coupling for the positive-mass-squared 
case.  An apparent discrepancy with equal-time calculations of
the critical coupling is resolved by an appropriate mass
renormalization.

\end{abstract}

%
\pacs{12.38.Lg, 11.15.Tk, 11.10.Ef
}

\maketitle

\section{Introduction}
\label{sec:Introduction}

Although two-dimensional $\phi^4$ theory has a simple Lagrangian, the
spectrum of the theory has some very interesting behavior.  Even without
the introduction of a negative bare-mass squared, the theory exhibits
symmetry breaking at sufficiently strong coupling~\cite{Chang}, signaled by a
degeneracy of the lowest massive states with the vacuum.  Here we
consider a light-front Hamiltonian calculation of the spectrum
at and below this critical coupling and compare with previous calculations
in both light-front~\cite{VaryHari} and equal-time 
quantization~\cite{RychkovVitale,LeeSalwen,Sugihara,SchaichLoinaz,Bosetti,Milsted}.
In particular, we explain why the two quantizations yield different
results for the critical value of the bare dimensionless coupling.

Light-front quantization~\cite{DLCQreview} is in general a convenient approach to 
the nonperturbative solution of quantum field theories~\cite{Vary,ArbGauge},
and provides an alternative to lattice~\cite{Lattice} and Dyson--Schwinger methods~\cite{DSE}.  
Light-front coordinates~\cite{Dirac} offer a clean separation between external
and internal momenta, and the quantization can keep the vacuum trivial,
so that it does not mix with the massive states.  The wave functions
of a Fock-state expansion are then well defined and lay an intuitive
foundation for calculation of observables directly in terms of matrix 
elements of operators.  Moreover, the formulation
is in Minkowski space rather than the Euclidean space of lattice gauge theory
and Dyson--Schwinger equations.

In two dimensions, light-front quantization uses
a time coordinate $x^+\equiv t+z$ and a spatial coordinate
$x^-\equiv t-z$.  The conjugate momentum variables are $p^-\equiv E-p_z$
and $p^+\equiv E+p_z$, respectively.  The fundamental eigenvalue problem
for an eigenstate of mass $M$ is 
$\Pminus|\psi(P^+)\rangle=\frac{M^2}{P^+}|\psi(P^+)\rangle$, where
$\Pminus$ is the light-front Hamiltonian and $P^+$ is the total
light-front momentum of the state.  We solve this eigenvalue
problem by expanding $|\psi(P^+)\rangle$ in a Fock basis
of momentum and particle-number eigenstates and then expanding
the Fock-space wave functions in terms of fully symmetric,
multivariate polynomials~\cite{GenSymPolys}.  The problem is
made finite by truncation in both the Fock-space and
polynomial basis sets.

The use of a polynomial basis for the wave functions has significant
advantages over the more common discretized light-cone quantization (DLCQ)
approach~\cite{PauliBrodsky,DLCQreview}.  One purpose of the present
work is to illustrate this.  In DLCQ, which relies on a trapezoidal
approximation to integral operators, endpoint corrections associated
with zero modes~\cite{ZeroModes} are normally dropped,\footnote{The 
standard DLCQ approach of neglecting zero modes can be modified to 
include them~\protect\cite{DLCQzeromodes}.} which delays convergence,
whereas a basis-function method can be tuned to keep the correct
endpoint behavior of the wave functions.  Also, the discretization
grid of DLCQ forces a particular allocation of computational
resources to each Fock sector, without regard to the importance
of one sector over another; a basis-function approach allows
the allocation to be adjusted sector by sector, to optimize 
computation resources with respect to convergence.  The particular
polynomial basis that we use~\cite{GenSymPolys} is specifically 
symmetric with respect to interchange of the identical bosons, so 
that no explicit symmetrization is required.

The remainder of the paper is structured as follows.  Section~\ref{sec:EVProblem}
describes the eigenvalue problem that we solve, with details
of the coupled integral equations for the Fock-state wave functions.
In Sec.~\ref{sec:Mass} we discuss the difference in mass
renormalizations for light-front and equal-time quantization
and provide a scheme for calculation.  Our results are presented
and discussed in Sec.~\ref{sec:Results}, with a brief summary
provided in Sec.~\ref{sec:Summary}.  Details of the numerical
calculation are given in an Appendix.

\section{Light-front eigenvalue problem}
\label{sec:EVProblem}

From the Lagrangian for two-dimensional $\phi^4$ theory
\be
{\cal L}=\frac12(\partial_\mu\phi)^2-\frac12\mu^2\phi^2-\frac{\lambda}{4!}\phi^4,
\ee
where $\mu$ is the mass of the boson and $\lambda$ is the coupling constant,
the light-front Hamiltonian density is found to be
\be
{\cal H}=\frac12 \mu^2 \phi^2+\frac{\lambda}{4!}\phi^4.
\ee
The mode expansion for the field at zero light-front time is
\be \label{eq:mode}
\phi(x^+=0,x^-)=\int \frac{dp}{\sqrt{4\pi p}}
   \left\{ a(p)e^{-ipx^-/2} + a^\dagger(p)e^{ipx^-/2}\right\},
\ee
where for convenience we have dropped the $+$ superscript
and will from here on write light-front momenta such as
$p^+$ as just $p$.
The creation operator $a^\dagger(p)$ satisfies the commutation
relation
\be
[a(p),a^\dagger(p')]=\delta(p-p'),
\ee
and builds $m$-constituent Fock states from the Fock vacuum $|0\rangle$ in the form
\be
|y_iP;P,m\rangle=\frac{1}{\sqrt{m!}}\prod_{i=1}^m a^\dagger(y_iP)|0\rangle.
\ee
Here $y_i\equiv p_i/P$ is the longitudinal momentum fraction for the $i$th constituent.

The light-front Hamiltonian is 
$\Pminus=\Pminus_{11}+\Pminus_{22}+\Pminus_{13}+\Pminus_{31}$,
with
\bea \label{eq:Pminus11}
\Pminus_{11}&=&\int dp \frac{\mu^2}{p} a^\dagger(p)a(p),  \\
\label{eq:Pminus22}
\Pminus_{22}&=&\frac{\lambda}{4}\int\frac{dp_1 dp_2}{4\pi\sqrt{p_1p_2}}
       \int\frac{dp'_1 dp'_2}{\sqrt{p'_1 p'_2}} 
       \delta(p_1 + p_2-p'_1-p'_2) \\
 && \rule{2in}{0mm} \times a^\dagger(p_1) a^\dagger(p_2) a(p'_1) a(p'_2),
   \nonumber \\
\label{eq:Pminus13}
\Pminus_{13}&=&\frac{\lambda}{6}\int \frac{dp_1dp_2dp_3}
                              {4\pi \sqrt{p_1p_2p_3(p_1+p_2+p_3)}} 
     a^\dagger(p_1+p_2+p_3)a(p_1)a(p_2)a(p_3), \\
\label{eq:Pminus31}
\Pminus_{31}&=&\frac{\lambda}{6}\int \frac{dp_1dp_2dp_3}
                              {4\pi \sqrt{p_1p_2p_3(p_1+p_2+p_3)}} 
      a^\dagger(p_1)a^\dagger(p_2)a^\dagger(p_3)a(p_1+p_2+p_3).
\eea
The subscripts indicate the number of creation and annihilation operators
in each term.

The Fock-state expansion of an eigenstate can be written
\be \label{eq:FSexpansion}
|\psi(P)\rangle=\sum_m P^{\frac{m-1}{2}}\int\prod_i^m dy_i 
       \delta(1-\sum_i^m y_i)\psi_m(y_i)|y_iP;P,m\rangle,
\ee
where $\psi_m$ is the wave function for $m$ constituents.
Because the terms of $\Pminus$ change particle
number by zero or by two, the eigenstates can be separated 
according to the oddness or evenness of the number of constituents.
Therefore, the first sum in (\ref{eq:FSexpansion}) is restricted to
odd or even $m$.  We will consider only the lowest mass eigenstate
in each case, though the methods allow for calculation of higher
states.

The light-front Hamiltonian eigenvalue problem 
${\cal P}^-|\psi(P)\rangle=\frac{M^2}{P}|\psi(P)\rangle$
reduces to a coupled set of integral equations for the
Fock-state wave functions:
\bea
\lefteqn{m\frac{\mu^2}{y_1 P}\psi_m(y_i)
+\frac{\lambda}{4\pi P}\frac{m(m-1)}{4\sqrt{y_1y_2}}
        \int\frac{dx_1 dx_2 }{\sqrt{x_1 x_2}}\delta(y_1+y_2-x_1-x_2) \psi_m(x_1,x_2,y_3,\ldots,y_m)}&& 
        \nonumber \\
&& +\frac{\lambda}{4\pi P}\frac{m}{6}\sqrt{(m+2)(m+1)}\int \frac{dx_1 dx_2 dx_3}{\sqrt{y_1 x_1 x_2 x_3}}
        \delta(y_1-x_1-x_2-x_3)\psi_{m+2}(x_1,x_2,x_3,y_2,\ldots,y_m) \nonumber \\
&& +\frac{\lambda}{4\pi P}\frac{m-2}{6}\frac{\sqrt{m(m-1)}}{\sqrt{y_1y_2y_3(y_1+y_2+y_3)}}
          \psi_{m-2}(y_1+y_2+y_3,y_4,\ldots,y_m)=\frac{M^2}{P}\psi_m(y_i).
\eea
We have used the symmetry of $\psi_m$ to collect exchanged momenta in
the leading arguments of the function, with appropriate $m$-dependent
factors in front of each term.  The equations are simplified further
by the introduction of a dimensionless coupling
\be
g\equiv\frac{\lambda}{4\pi\mu^2}
\ee
and by multiplying the set by $P/\mu^2$, to obtain
\bea \label{eq:coupledsystem}
\lefteqn{\frac{m}{y_1}\psi_m(y_i)
+\frac{g}{4}\frac{m(m-1)}{\sqrt{y_1y_2}}
        \int\frac{dx_1 dx_2 }{\sqrt{x_1 x_2}}\delta(y_1+y_2-x_1-x_2) \psi_m(x_1,x_2,y_3,\ldots,y_m)}&& 
        \nonumber \\
&& +\frac{g}{6}m\sqrt{(m+2)(m+1)}\int \frac{dx_1 dx_2 dx_3}{\sqrt{y_1 x_1 x_2 x_3}}
        \delta(y_1-x_1-x_2-x_3)\psi_{m+2}(x_1,x_2,x_3,y_2,\ldots,y_m) \nonumber \\
&& +\frac{g}{6}\frac{(m-2)\sqrt{m(m-1)}}{\sqrt{y_1y_2y_3(y_1+y_2+y_3)}}
          \psi_{m-2}(y_1+y_2+y_3,y_4,\ldots,y_m)=\frac{M^2}{\mu^2}\psi_m(y_i).
\eea
It is this set of equations that we solve numerically, as described in the
Appendix.  Our approach takes advantage of the new set of multivariate
polynomials that is fully symmetric on the hypersurface $\sum_i y_i=1$
defined by momentum conservation~\cite{GenSymPolys}.  This allows independent 
tuning of resolutions in each Fock sector, so that, unlike discrete
light-cone quantization (DLCQ)~\cite{PauliBrodsky,DLCQreview}, sectors 
with lower net probability need not overtax computational resources.
Also, within each Fock sector, the use of a polynomial basis has improved
convergence compared to DLCQ~\cite{GenSymPolys}, at least partly because
DLCQ misses contributions from zero modes~\cite{ZeroModes}
associated with integrable singularities at $y_i=0$.

In addition to calculation of the spectrum, it is possible to calculate the expectation
value for the field $\phi$ when the odd and even states are degenerate.
At degeneracy, the two states mix, and the expectation value for $\phi$
comes from cross terms, the matrix element between the odd and even
eigenstates.  Let $|\widetilde\psi(P')\rangle$ be the state with an
odd number of constituents, and $|\psi(P)\rangle$ be the state for
an even number.  At degeneracy, the eigenstate can be a linear
combination of these, and the desired matrix element for the
field is
\bea
\lefteqn{\langle\widetilde\psi(P')|\phi(0,x^-)|\psi(P)\rangle=
\sum_m \frac{P^{\prime\,m/2-1}}{P^{(m-1)/2}}\int\prod_j^{m+1} dy'_j\delta(1-\sum_j y'_j)
     \frac{\sqrt{m+1}}{\sqrt{4\pi y'_1 P'}}} && \nonumber \\
 && \rule{1in}{0mm} \times \delta(y'_1+P/P'-1)e^{i(P'-P)x^-/2} \widetilde\psi_{m+1}(y'_j)\psi_m(y'_2,...,y'_{m+1}) \\
&& \rule{0.5in}{0mm}+\sum_m \frac{P^{m/2-1}}{P^{\prime\,(m-1)/2}}\int\prod_j^{m+1} dy_j\delta(1-\sum_j y_j)
     \frac{\sqrt{m+1}}{\sqrt{4\pi y_1 P}}\delta(y_1+P'/P-1) \nonumber \\
 && \rule{1in}{0mm} \times    e^{-i(P'-P)x^-/2} \widetilde\psi_m(y_2,\ldots,y_{m+1})\psi_{m+1}(y_j), \nonumber
\eea
where again we have taken advantage of the wave-function symmetry to
arrange for all but the first constituent to be spectators.  In the
limit $P'\rightarrow P$, this expression reduces to
\bea \label{eq:vev}
\lefteqn{\langle\widetilde\psi(P)|\phi(0,x^-)|\psi(P)\rangle=
\frac12\sum_m\frac{\sqrt{m+1}}{\sqrt{4\pi P}}\int \prod_{i=2}^{m+1} dy_i \delta(1-\sum_i y_i)}&& \\
&& \rule{0.5in}{0mm} \times \left[\widetilde\phi_{m+1}(y_2,\ldots,y_{m+1})\psi_m(y_2,\ldots,y_{m+1})
+\widetilde\psi_m(y_2,\ldots,y_{m+1})\phi_{m+1}(y_2,\ldots,y_{m+1}) \right], \nonumber
\eea
with
\be
\phi_{m+1}(y_2,\ldots,y_{m+1})\equiv \lim_{y_1\rightarrow0}\frac{1}{\sqrt{y_1}}\psi_{m+1}(y_1,y_2,\ldots,y_m)
\ee
and $\widetilde\phi$ defined analogously.
Thus the expectation value depends upon zero modes~\cite{ZeroModes}.
In our basis function approach, the zero-momentum limit can be taken
explicitly.

\section{Mass renormalization}
\label{sec:Mass}

One of the advantages of light-front quantization is the absence of
vacuum-to-vacuum graphs~\cite{DLCQreview}.  However, to compare 
results found in equal-time quantization at equivalent values
of the bare parameters in the Lagrangian, this absence must be
taken into account.  In particular, the bare mass in $\phi^4_{1+1}$ theory
is renormalized by tadpole contributions in equal-time quantization
but not in light-front quantization, and the two different masses
are related by~\cite{SineGordon}
\be
\mu_{\rm LF}^2=\mu_{\rm ET}^2
   +\lambda\left[\langle 0|:\frac{\phi^2}{2}:|0\rangle
       -\langle 0|:\frac{\phi^2}{2}:|0\rangle_{\rm free}\right].
\ee
The vacuum expectation values (vev) of $\phi^2$ resum the tadpole
contributions; the subscript {\em free} indicates the vev with
$\lambda=0$.  This distinction between bare masses in the
two quantizations implies that the dimensionless coupling $g=\lambda/4\pi\mu^2$
is also not the same.  Estimates of the critical coupling must
then be adjusted for the difference if they are to be compared.

Of course, if one compares results only for physical quantities,
the two quantizations should match immediately.  However, this
is not straightforward in the case of the critical coupling, where 
the physical mass scale goes to zero.

With the tadpole contribution re-expressed as a vev, we
can calculate the contribution in light-front quantization
and avoid doing a second, equal-time calculation.  The
vev is regulated by point splitting, and a sum over a
complete set states is introduced, to obtain
\be
\langle 0|:\frac{\phi^2}{2}:|0\rangle\rightarrow 
   \frac12\langle 0|\phi(\epsilon^+,\epsilon^-)\int_0^\infty dP\sum_n|\psi_n(P)\rangle
      \langle\psi_n(P)|\phi(0,0)|0\rangle.
\ee
Each mass eigenstate $|\psi_n(P)\rangle$ is expanded in terms of Fock states
and wave functions, just as in (\ref{eq:FSexpansion}), with
the Fock wave functions $\psi_{nm}(y_i)$ now defined with an additional index $n$
for the particular eigenstate.  Because the $\phi$ field changes particle number
by one, only one-particle Fock states will contribute to the sum over $n$,
with amplitude $\psi_{n1}$.  For the free case, the only contribution
to the sum is the one-particle state $a^\dagger(P)|0\rangle$.

The individual matrix elements are readily calculated.  At $x^+=0$ the
field is given by (\ref{eq:mode}), and the matrix element is
\be
\langle\psi_n(P)|\phi(0,0)|0\rangle
  =\langle 0|\psi_{n1}^*a(P)\int\frac{dp}{\sqrt{4\pi p}}a^\dagger(p)|0\rangle
  =\frac{\psi_{n1}^*}{\sqrt{4\pi P}}
\ee
At $x^+=\epsilon^+$, the field is
\be
\phi(\epsilon^+,\epsilon^-)=e^{i\Pminus\epsilon^+/2}\phi(0,\epsilon^-)e^{-i\Pminus\epsilon^+/2}.
\ee
The matrix element is
\bea
\langle 0|\phi(\epsilon^+,\epsilon^-)|\psi_n(P)\rangle
&=&\langle 0|e^{i0\epsilon^+}\int\frac{dp}{\sqrt{4\pi p}}a(p)
  e^{-ip\epsilon^-/2}e^{-iM_n^2\epsilon^+/2P}\psi_{n1}a^\dagger(P)|0\rangle \nonumber \\
  &=&\frac{\psi_{n1}}{\sqrt{4\pi P}}e^{-i(P\epsilon^- +M_n^2\epsilon^+/P)/2},
\eea
with $M_n$ the mass of the $n$th state.  The corresponding matrix elements
for the free case are
\be
\langle0|a(P)\phi(0,0)|0\rangle
  =\langle 0|a(P)\int\frac{dp}{\sqrt{4\pi p}}a^\dagger(p)|0\rangle
  =\frac{1}{\sqrt{4\pi P}}
\ee
and
\bea
\langle 0|\phi(\epsilon^+,\epsilon^-)a^\dagger(P)|0\rangle
&=&\langle 0|e^{i0\epsilon^+}\int\frac{dp}{\sqrt{4\pi p}}a(p)
  e^{-ip\epsilon^-/2}e^{-i\mu^2\epsilon^+/2P}a^\dagger(P)|0\rangle \nonumber \\
  &=&\frac{1}{\sqrt{4\pi P}}e^{-i(P\epsilon^- +\mu^2\epsilon^+/P)/2}.
\eea

The combination of these matrix elements yields the two vev's:
\be
\langle 0|:\frac{\phi^2}{2}:|0\rangle=\frac12\sum_n\int_0^\infty dP \frac{|\psi_{n1}|^2}{4\pi P}
       e^{-i(P\epsilon^- +M_n^2\epsilon^+/P)/2}
\ee
and
\be
\langle 0|:\frac{\phi^2}{2}:|0\rangle_{\rm free}=\frac12\int_0^\infty dP \frac{1}{4\pi P}
       e^{-i(P\epsilon^- +\mu^2\epsilon^+/P)/2}.
\ee
The completeness of the eigenstates allows the introduction of the sum $1=\sum_n|\psi_{n1}|^2$
into the free vev, so that the difference can be written
\be
\langle 0|:\frac{\phi^2}{2}:|0\rangle-\langle 0|:\frac{\phi^2}{2}:|0\rangle_{\rm free}
=\sum_n \frac{|\psi_{n1}|^2}{8\pi}\int_0^\infty \frac{dP}{P} e^{-iP\epsilon^-/2}
\left[e^{-i\frac{M_n^2\epsilon^+}{2P}}-e^{-i\frac{\mu^2\epsilon^+}{2P}}\right].
\ee
With the change of variable $P=2z\epsilon^+$ and the introduction of a convergence
factor $e^{-\eta z}$, this expression becomes
\be
\langle 0|:\frac{\phi^2}{2}:|0\rangle-\langle 0|:\frac{\phi^2}{2}:|0\rangle_{\rm free}
=\sum_n \frac{|\psi_{n1}|^2}{8\pi}\int_0^\infty \frac{dz}{z} e^{iz(-\epsilon^2+i\eta)}
\left[e^{-i\frac{M_n^2}{4z}}-e^{-i\frac{\mu^2}{4z}}\right].
\ee
Each term is an integral representation~\cite{GradshteynRyzhik} of the modified 
Bessel function $K_0$:
\be
\int_0^\infty \frac{dx}{x}\exp\left(i\frac{\alpha}{2}\left[x-\frac{\beta^2}{x}\right]\right)=2K_0(\alpha\beta),
\ee
with positive imaginary parts for $\alpha$ and $\alpha\beta^2$.
The difference of vev's then becomes
\be
\langle 0|:\frac{\phi^2}{2}:|0\rangle-\langle 0|:\frac{\phi^2}{2}:|0\rangle_{\rm free}
=\sum_n \frac{|\psi_{n1}|^2}{4\pi}\left[K_0(M_n\sqrt{-\epsilon^2+i\eta})
                 -K_0(\mu\sqrt{-\epsilon^2+i\eta})\right].
\ee
As $-\epsilon^2+i\eta$ goes to zero, the only contribution from $K_0$ is a simple
logarithm, i.e. $K_0(z)\rightarrow-\ln(z/2)-\gamma$, leaving
\be \label{eq:Delta}
\langle 0|:\frac{\phi^2}{2}:|0\rangle-\langle 0|:\frac{\phi^2}{2}:|0\rangle_{\rm free}
=-\sum_n \frac{|\psi_{n1}|^2}{4\pi}\ln\frac{M_n}{\mu_{\rm LF}}\equiv -\Delta/4\pi,
\ee
with $\mu$ written as $\mu_{\rm LF}$ to emphasize that it is the light-front bare mass.
Within the context of our numerical calculation, the terms of the sum can be computed by
fully diagonalizing the matrix representation of the Hamiltonian $\Pminus$.

The bare masses in the two quantizations are then related by
\be 
\mu_{\rm LF}^2=\mu_{\rm ET}^2-\frac{\lambda}{4\pi}\Delta \;\;
\mbox{or} \;\;
\frac{\mu_{\rm ET}^2}{\mu_{\rm LF}^2}=1+g_{\rm LF}\Delta.
\ee
The bare-mass ratio is what connects the dimensionless couplings and masses
obtained in the two quantizations:
\be \label{eq:couplingratio}
g_{\rm ET}=\frac{g_{\rm LF}}{\mu_{\rm ET}^2/\mu_{\rm LF}^2}=\frac{g_{\rm LF}}{(1+g_{\rm LF}\Delta)}\;\;
\mbox{and} \;\;
\frac{M^2}{\mu_{\rm ET}^2}=\frac{M^2/\mu_{\rm LF}^2}{\mu_{\rm ET}^2/\mu_{\rm LF}^2}
      =\frac{1}{1+g_{\rm LF}\Delta}\frac{M^2}{\mu_{\rm LF}^2},
\ee
which we can use to compare the values obtained.

\section{Results and discussion}
\label{sec:Results}

\begin{figure}[ht]
\vspace{0.2in}
\centerline{\includegraphics[width=15cm]{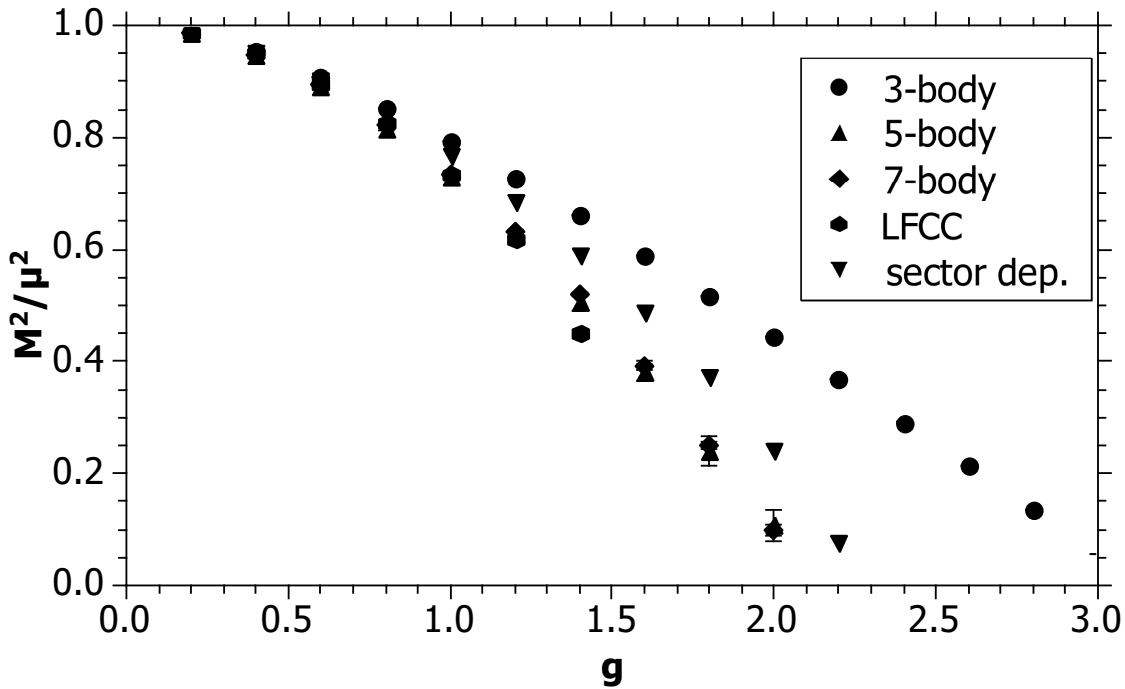}}
\caption{\label{fig:extrap-odd}
Lowest mass eigenvalue for odd numbers of constituents.
Results are shown for different Fock-space truncations
to three, five, and seven constituents.  Also plotted
are results for the leading light-front coupled-cluster (LFCC)
approximation and for a sector-dependent modification of
the three-body truncation.  The errors are estimated
from extrapolations in polynomial basis size.
}
\end{figure}

\begin{figure}[ht]
\vspace{0.2in}
\centerline{\includegraphics[width=15cm]{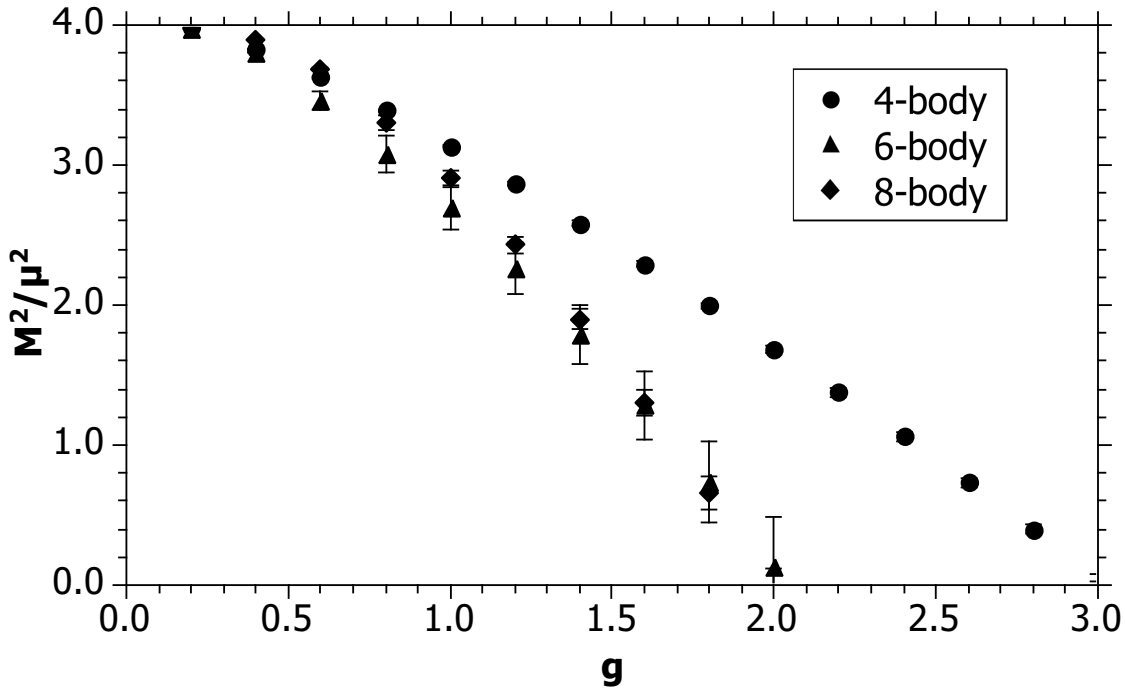}}
\caption{\label{fig:extrap-even} 
Same as Fig.~\ref{fig:extrap-odd} but for even numbers
of constituents, with Fock-space truncations at four,
six and eight.
}
\end{figure}

With the numerical methods discussed in the Appendix, we have solved the
eigenvalue problem for the lowest odd and even states.  The mass values
for different Fock-space truncations are shown in Figs.~\ref{fig:extrap-odd}
and \ref{fig:extrap-even}.  The error bars are estimated based on
the extrapolations in basis size.  The seven/eight-body truncations yield 
results that are the same as the five/six-body results, to within the 
error estimates, which means that convergence in the Fock-state
expansion has been achieved.  

In the odd case, we also show results from
the leading light-front coupled-cluster (LFCC) 
approximation~\cite{LFCC,LFCCphi4}\footnote{There are sign errors in Eq.~(B4)
of \protect\cite{LFCCphi4}.  The signs of the $A$ and $B$ terms should be
reversed; however, the computations were done with the correct signs.}
and a modification of the three-body truncation that includes sector-dependent
bare masses~\cite{Wilson,hb,Karmanov,SecDep}.  The LFCC calculation includes 
a partial summation over all
higher Fock states.  The sector-dependent calculation uses the physical
mass in the upper Fock sector, where there can be no self-energy correction.
Both of these alternatives require solution of a three-body problem
and yield results much better than the simple three-body truncation,
with the LFCC approximation doing much better than the sector-dependent
approach.

\begin{figure}[ht]
\vspace{0.2in}
\centerline{\includegraphics[width=15cm]{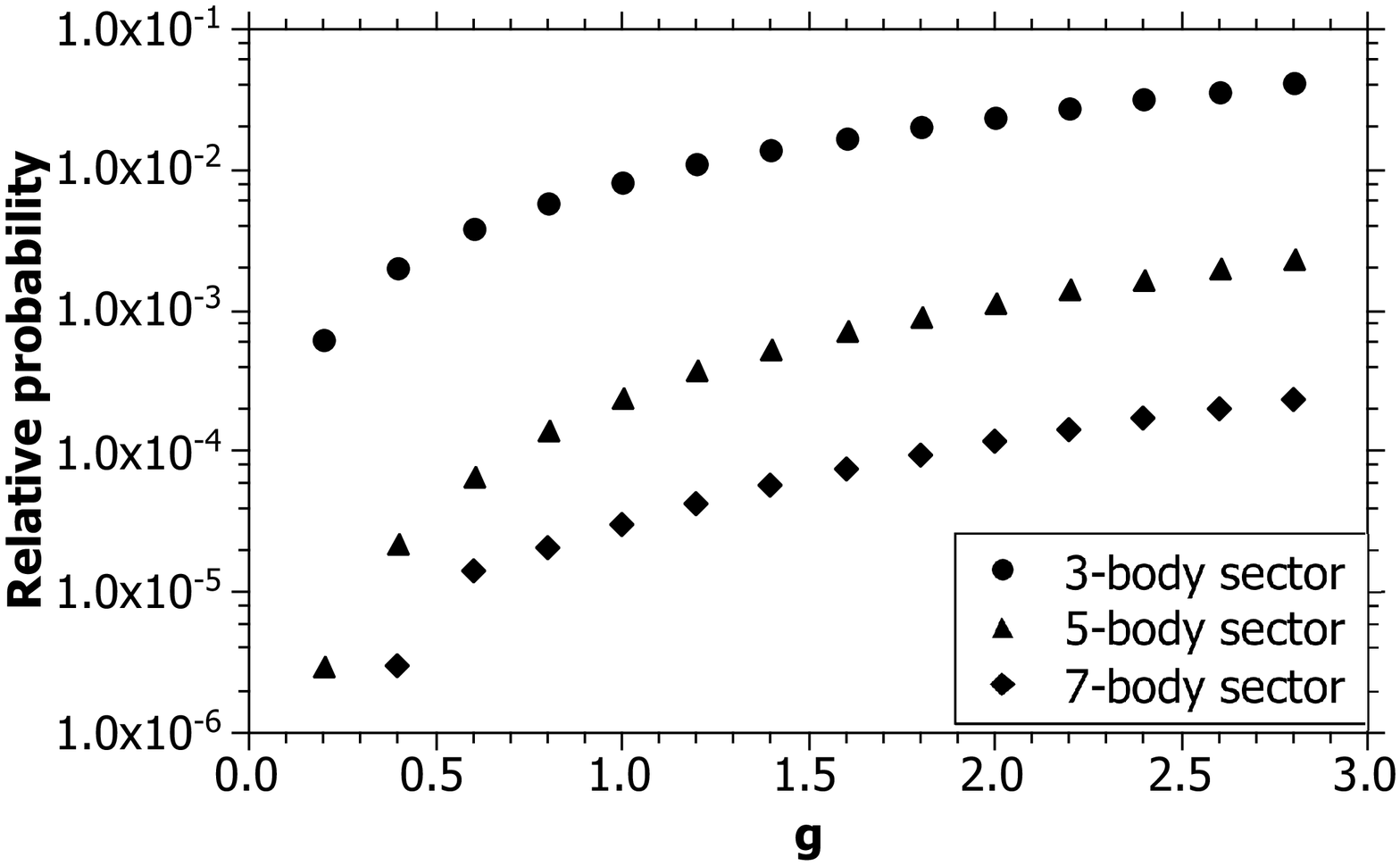}}
\caption{\label{fig:relprob} 
Relative Fock-sector probabilities for the lowest mass eigenstate with
odd numbers of constituents.
}
\end{figure}

The relative Fock-sector probabilities for the odd case are plotted 
in Fig.~\ref{fig:relprob}.  These ratios are computed as
\be
R_m\equiv\frac{1}{|\psi_1|^2}\int\left[\prod_{i=1}^{m-1}dy_i\right]|\psi_m(y_i)|^2
   =\frac{1}{|\psi_1|^2}\sum_{ni}|c_{ni}^{(m)}|^2,
\ee
where the last expression is in terms of the basis-function 
expansion coefficients, as defined in (\ref{eq:expansion}).
These ratios show that the probability for each
Fock sector decreases by an order of magnitude when the number of
constituents goes up by two.

The apparent convergence of the Fock-state expansion is somewhat deceptive.
With the bare mass fixed as the same in all Fock sectors, the higher
Fock sectors are suppressed by the large invariant mass of each
Fock state, which is of order $m\mu$ for the sector with $m$
constituents.  For weak to moderate couplings this is not a 
particular concern, but for strong coupling, approaching the
critical value, one expects much larger contributions from
higher Fock states.  This would be best modeled by sector-dependent
masses.\footnote{An alternative is the LFCC method~\protect\cite{LFCCphi4},
which automatically uses the physical mass for the kinetic
energy contributions to the wave equations.}  However, to
use sector-dependent masses requires renormalization to
physical observables, which would greatly complicate any
comparison with the published results for equal-time 
quantization.  Therefore, for purposes of the the present
work, we retain a fixed bare mass.

\begin{figure}[ht]
\vspace{0.2in}
\centerline{\includegraphics[width=15cm]{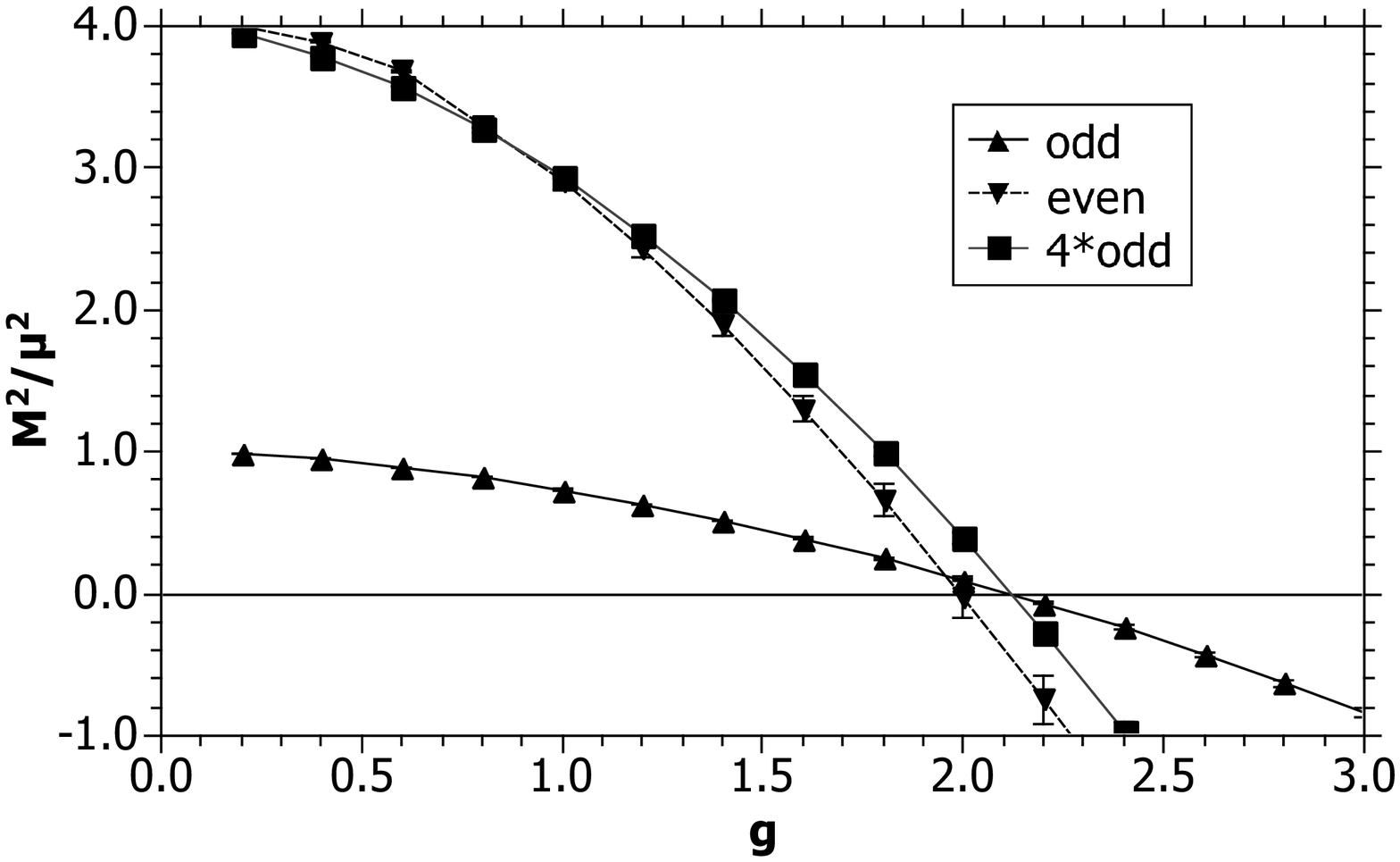}}
\caption{\label{fig:critcoup} 
The lowest masses for the odd and even cases, as used
to estimate the critical coupling, including a plot
of the threshold for two-particle states at four 
times the mass-squared of the odd case.
}
\end{figure}

Based on the results for the masses as a function of the
coupling, we can estimate a critical coupling as the value
at which the lowest masses reach zero.  This intersection
is illustrated in Fig.~\ref{fig:critcoup}, where the
lowest mass-squared values are plotted as well as four
times the odd-eigenstate mass squared.  Because there 
are no bound states in this theory the lowest even-eigenstate
mass should be equal to this; the difference in the plot
is another measure of the numerical and truncation errors.
From this plot, we estimate the critical value of the
dimensionless coupling to be 2.1$\pm$0.05.  For comparison,
we list in Table~\ref{tab:critcoup} this and values
from other computations, as gathered in \cite{RychkovVitale}; 
however, because the definitions of dimensionless couplings vary, 
the table uses the definition $\bar{g}\equiv\lambda/24\mu^2$,
which is just $\frac{\pi}{6}g$.

\begin{table}[ht]
\caption{\label{tab:critcoup}
Comparison of critical coupling values, adapted 
from~\protect\cite{RychkovVitale}, with a slightly
different definition of the dimensionless
coupling $\bar{g}=\frac{\pi}{6}g$.
The first two values were computed
in light-front quantization and the remainder in equal-time
quantization; the results are comparable but only after the
different mass renormalizations are taken into account,
as discussed in the text.}
\begin{center}
\begin{tabular}{lll}
\hline \hline
Method &  $\bar{g}_c$ & Reported by \\
\hline
Light-front symmetric polynomials & $1.1\pm0.03$ & this work \\
DLCQ  & 1.38 & Harindranath \& Vary~\protect\cite{VaryHari} \\
Quasi-sparse eigenvector & 2.5 & 
     Lee \& Salwen~\protect\cite{LeeSalwen} \\
Density matrix renormalization group & 2.4954(4) & 
     Sugihara~\protect\cite{Sugihara} \\
Lattice Monte Carlo & 2.70$\left\{\begin{array}{l} +0.025 \\ -0.013\end{array}\right.$ & 
     Schaich \& Loinaz~\protect\cite{SchaichLoinaz} \\
                    & $2.79\pm0.02$ & Bosetti {\em et al.}~\protect\cite{Bosetti} \\
Uniform matrix product & 2.766(5) & 
     Milsted {\em et al.}~\protect\cite{Milsted} \\
Renormalized Hamiltonian truncation & 2.97(14) & 
     Rychkov \& Vitale~\protect\cite{RychkovVitale} \\
\hline \hline
\end{tabular}
\end{center}
\end{table}

\begin{figure}[ht]
\vspace{0.2in}
\centerline{\includegraphics[width=15cm]{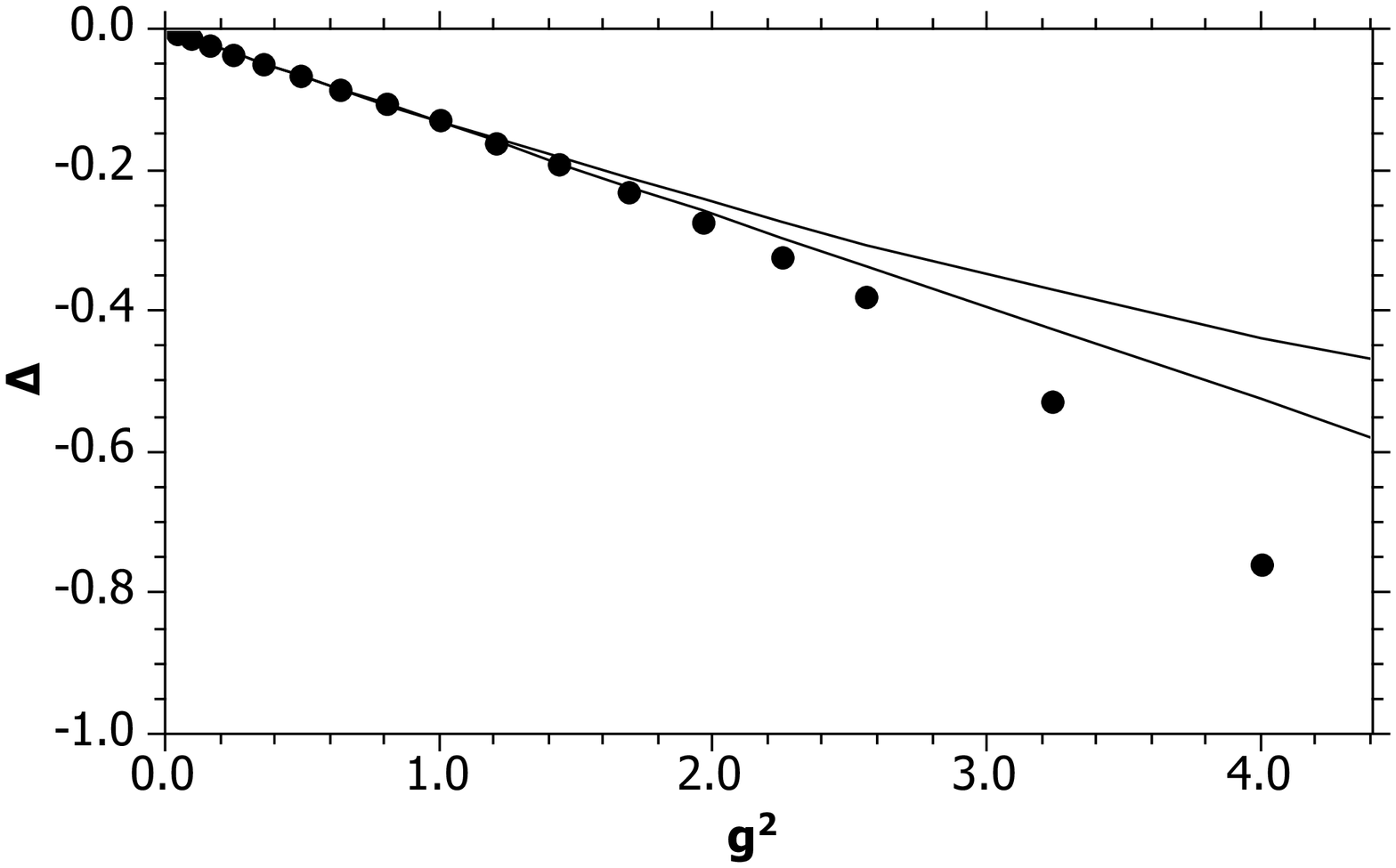}}
\caption{\label{fig:shift}
The renormalization shift $\Delta$, defined in Eq.~(\ref{eq:Delta})
of the text, as a function of the square of the dimensionless coupling $g$.
The points displayed are obtained as extrapolations in the polynomial
basis size.  The lines are linear and quadratic fits to shifts below 
$g=1$, extrapolated to the region of the critical coupling.}
\end{figure}

The results listed in Table~\ref{tab:critcoup}
imply a systematic difference between equal-time and light-front values
for the critical coupling, which is exactly what should be expected,
based on the difference in mass renormalizations discussed in
Sec.~\ref{sec:Mass}.  To quantify this difference, we extended the
diagonalization of the Hamiltonian matrix to include the entire
spectrum and computed the shift $\Delta$, defined in (\ref{eq:Delta}).
The results are plotted in Fig.~\ref{fig:shift}, along with 
extrapolations of fits to the shifts for coupling values below 1.

\begin{figure}[ht]
\vspace{0.2in}
\centerline{\includegraphics[width=15cm]{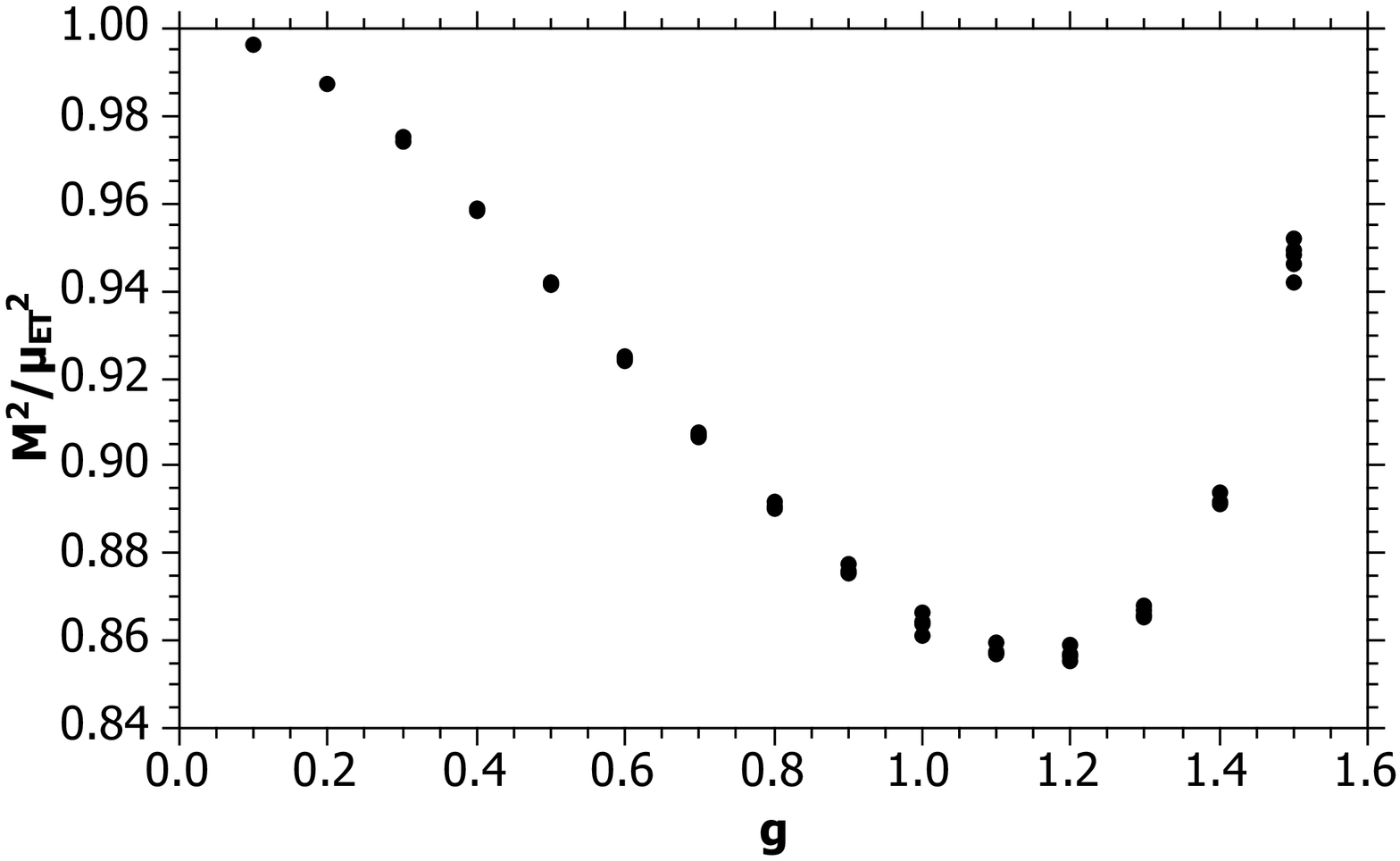}}
\caption{\label{fig:ETmass}
Lowest equal-time mass eigenvalues for odd numbers of constituents
plotted versus the dimensionless light-front coupling $g$.
Each mass value is obtained from the light-front masses and 
the computed shift $\Delta$ according to Eq.~(\ref{eq:couplingratio})
of the text.  Different points at the same $g$ value correspond to
different truncations of the polynomial basis size.
}
\end{figure}

Higher coupling values are not used in the fits because the lack of 
sector-dependent mass renormalization does not allow a 
reasonable approximation to the wave functions.  The one-particle
sector should become less and less probable for the lowest eigenstate,
as the critical coupling is approached and as its mass approaches
zero; instead, the one-particle probability remains finite and
the product $|\psi_{11}|\ln(M_1^2)$ diverges.  To judge the
value of the coupling where this effect becomes noticeable,
we studied the behavior of the dimensionless mass $M^2/\mu_{\rm ET}^2$,
as predicted by (\ref{eq:couplingratio}), as a function of
the coupling, which we plot in Fig.~\ref{fig:ETmass}.  For 
coupling values above 1, the mass begins to increase rather
than decrease; this incorrect behavior is the precursor of
the divergence at the critical coupling; we also see that
the convergence with respect to the polynomial basis becomes
somewhat worse at these larger coupling values.

The estimated value of the shift at the critical 
coupling 2.1, based on the two extrapolations,
is $\Delta(g=2.1)=-0.47\pm0.12$.  The value is 
from the higher-order extrapolation, with the
lower-order extrapolation used to indicate the
error.  From the latest equal-time value
for the critical coupling~\cite{RychkovVitale},
$g_{\rm ETc}=\frac{6}{\pi}2.97=5.67$, we extract
a shift of $(g_{\rm LFc}/g_{\rm ETc}-1)/g_{\rm LFc}=-0.30$,
which is consistent with the estimated value of the shift.

\section{Summary}
\label{sec:Summary}

We have developed a high-order method
for ($1+1$)-dimensional light-front theories 
that is distinct from DLCQ~\cite{PauliBrodsky,DLCQreview}.  It is based on fully symmetric
multivariate polynomials~\cite{GenSymPolys} that respect the momentum conservation constraint
and allows separate tuning of resolutions
in each Fock sector.  This method could be combined with transverse
discretization or basis functions for numerical solution of
($3+1$)-dimensional theories.

As an illustration, the method has been applied here
to $\phi^4_{1+1}$ theory.  The lowest mass eigenvalues
have been computed, as shown in Figs.~\ref{fig:extrap-odd} and \ref{fig:extrap-even};
they converge rapidly with respect to the Fock-space truncation.
The odd case includes comparison with the light-front coupled-cluster 
(LFCC) method~\cite{LFCC,LFCCphi4},
which indicates that the LFCC method combined with symmetric polynomials shows
promise for rapid convergence.
Either approach can also be applied to the negative-mass
squared case, where the symmetry breaking is explicit.

From the behavior of the mass eigenstates with respect to coupling strength,
as shown in Fig.~\ref{fig:critcoup}, we have extracted an estimate of
the critical coupling for $\phi^4_{1+1}$ theory with positive mass squared.
Above this coupling, the symmetry is broken.  With mass renormalization
properly taken into account, as discussed in Sec.~\ref{sec:Mass},
the value obtained ($g_c=2.1\pm0.05$ or $\bar{g}_c=1.1\pm0.03$) is 
comparable to values obtained in equal-time quantization.

The calculation can be improved by invocation of sector-dependent mass
renormalization~\cite{Wilson,hb,Karmanov,SecDep}, so that higher Fock
states can make a significant contribution as the coupling approaches
the critical value.  However, any comparison with equal-time quantization 
will then require the use of physical quantities as reference points,
rather than a direct comparison of bare couplings.

\acknowledgments
This work was supported in part by 
the Minnesota Supercomputing Institute through
grants of computing time and (for M.B.) supported in part by 
the US DOE under grant number FG03-95ER40965.

\appendix

\section{Numerical methods}  \label{sec:appendix}

The coupled system of equations for the Fock-state wave
functions are solved numerically using an expansion
in terms of fully symmetric polynomials~\cite{GenSymPolys}.
The coefficients of the expansion satisfy a matrix
eigenvalue problem, which is then diagonalized.
For the matrix problem to be finite, the Fock-state
expansion and the basis-function expansion are truncated.
We study the behavior of results as a function of 
the truncations and can make extrapolations from
simple fits.

\subsection{Matrix representation}

We expand the wave functions as
\be \label{eq:expansion}
\psi_m(y_i)=\sqrt{\prod_i y_i}\sum_{ni} c_{ni}^{(m)} P_{ni}^{(m)}(y_1,\ldots,y_m)
\ee
where the $P_{ni}^{(m)}$ are polynomials in the $m$ momentum fractions $y_i$
of order $n$ and the $c_{ni}^{(m)}$ are the expansion coefficients.  The
polynomials are fully symmetric with respect to interchange of the momenta;
the second subscript $i$ differentiates the various possibilities at a 
given order $n$.  For $m=2$ constituents there is only one possibility
at each order, but for $m>2$ there can be more than one.  However, the
number of linearly independent polynomials of a given order is restricted
by the momentum-conservation constraint $\sum_i y_i=1$.  

In \cite{GenSymPolys} we show that such polynomials can be written as
a product of powers of simpler polynomials, in the form
\be
P_{ni}^{(N)}=C_2^{n_2} C_3^{n_3}\cdots C_N^{n_N},
\ee
with the powers restricted by $n=\sum_j j n_j$.  Each different
way of decomposing $n$ into a sum of integers greater than 1
yields a different polynomial.  The $C_m$ are
sums of simple monomials $\prod_j^N y_j^{m_j}$ where $m_j$ is
zero or one and $\sum_j^N m_j=m$; the sum ranges over all
possible choices for the $m_j$, making each $C_m$ fully symmetric.
For example, given $N$ momentum variables, 
$C_2$ is $\sum_j^N \left(y_j\sum_{k>j}^N y_k\right)$,
$C_{N-1}$ is $\sum_j^N \prod_{k\neq j} y_k$,
and $C_N$ is $y_1y_2\cdots y_N$.  The first-order polynomial
$C_1=\sum_j y_j$ does not appear because the constraint reduces
it to a constant.

For the purposes of the present calculation, we do not explicitly
orthogonalize the polynomials.  An orthogonalization done
numerically via the Gram-Schmidt procedure or matrix diagonalization
methods results in too much round-off error for higher order
polynomials.  Analytic orthogonalization in exact arithmetic, 
as used in \cite{GenSymPolys}, avoids this but is unwieldy for 
high-order calculations with large numbers of constituents.
Here we use an implicit orthogonalization in the form of
a singular-value decomposition of the basis-function
overlap matrix, as discussed in the next section.

Given the expansion of the wave functions, the coupled system
of equations (\ref{eq:coupledsystem}) reduces to a set of
matrix equations
\be \label{eq:matrixequations}
\sum_{n'i'}\left[T^{(m)}_{ni,n'i'}+g V^{(m,m)}_{ni,n'i'}\right]c^{(m)}_{n'i'}
   +g \sum_{n'i'} V^{(m,m+2)}_{ni,n'i'} c^{(m+2)}_{n'i'}
   +g \sum_{n'i'} V^{(m,m-2)}_{ni,n'i'} c^{(m-2)}_{n'i'}
   =\frac{M^2}{\mu^2}\sum_{n'i'}B^{(m)}_{ni,n'i'}c_{n'i'}^{(m)},
\ee
where the kinetic-energy matrix is
\be
T^{(m)}_{ni,n'i'}=m\int\left(\prod_j dy_j \right)\delta(1-\sum_j y_j)\left(\prod_{j=2}^m y_j\right) P_{ni}^{(m)}(y_j)P_{n'i'}^{(m)}(y_j),
\ee
the potential-energy matrices are
\bea
V^{(m,m)}_{ni,n'i'}&=&\frac{g}{4}m(m-1)\int\left(\prod_j dy_j\right) \delta(1-\sum_j y_j) \\
  && \times  \int dx_1 dx_2 \delta(y_1+y_2-x_1-x_2)
      \left(\prod_{j=3}^m y_j\right) P_{ni}^{(m)}(y_j)P_{n'i'}^{(m)}(x_1,x_2,y_3,\ldots,y_m),
      \nonumber
\eea
\bea
\lefteqn{V^{(m,m+2)}_{ni,n'i'}=\frac{g}{6}m\sqrt{(m+2)(m+1)}\int\left(\prod_j dy_j\right) \delta(1-\sum_j y_j)}&& \\
  && \times 
    \int dx_1 dx_2 dx_3 \delta(y_1-x_1-x_2-x_3)
      \left(\prod_{j=2}^m y_j\right) P_{ni}^{(m)}(y_j)P_{n'i'}^{(m+2)}(x_1,x_2,x_3,y_2,\ldots,y_m),
      \nonumber
\eea
\bea
V^{(m,m-2)}_{ni,n'i'}&=&\frac{g}{6}(m-2)\sqrt{m(m-1)}\int\left(\prod_j dy_j\right)
        \delta(1-\sum_j y_j)  \\
  && \times 
      \left(\prod_{j=4}^m y_j \right)P_{ni}^{(m)}(y_j)P_{n'i'}^{(m-2)}(y_1+y_2+y_3,y_4,\ldots,y_m),
      \nonumber
\eea
and the basis-function overlap matrix is
\be
B^{(m)}_{ni,n'i'}=\int\left(\prod_j dy_j \right)\delta(1-\sum_j y_j)\left(\prod_j^m y_j\right) P_{ni}^{(m)}(y_j)P_{n'i'}^{(m)}(y_j).
\ee
All of the integrals can be done analytically in terms of the generalized beta function
\bea
B_m(m_1+1,m_2+1,\ldots,m_m+1)&=&\int \left(\prod_j dy_j)\right)\delta(1-\sum_j y_j)\left(\prod_j y_j^{m_j}\right) \\
   &=&\frac{m_1!m_2!\ldots m_m!}{(m_1+m_2+\cdots+m_m+m-1)!}, \nonumber
\eea
which can be computed recursively.  These matrix equations then define a symmetric
generalized eigenvalue problem, the solution of which is discussed in the next section.

The expectation value of the field can also be expressed in the given polynomial
basis and then computed directly from the expansion coefficients found in solving
the matrix problem.  Substitution of the expansion (\ref{eq:expansion}) into
the expression (\ref{eq:vev}) for the matrix element of the field yields
\bea
\lefteqn{\sqrt{4\pi P}\langle\widetilde\psi(P)|\phi(0,x^-)|\psi(P)\rangle=
\sum_m\frac{\sqrt{m+1}}{2}\sum_{n'i'}\sum_{ni}\int\left(\prod_{j=2}^{m+1} y_j dy_j\right)
  \delta(1-\sum_{j=2}^{m+1} y_j)}&& \nonumber  \\
 && \rule{0.5in}{0mm} \times P_{n'i'}^{(m+1)}(0,y_2,\ldots,y_{m+1})P_{ni}^{(m)}(y_2,\ldots,y_{m+1})
    \left\{\begin{array}{ll} \tilde c_{n'i'}^{(m+1)} c_{ni}^{(m)} & m\; {\rm even} \\ 
     c_{n'i'}^{(m+1)} \tilde c_{ni}^{(m)} & m\; {\rm odd}, \end{array}\right.
\eea
where the $\tilde c$ are the expansion coefficients for the odd eigenstate.

\subsection{Matrix diagonalization}

In principle, there are many ways to obtain the eigenvalues and 
eigenvectors of the generalized
problem (\ref{eq:matrixequations}), which we write here more compactly 
as $H\vec c=\xi B\vec c$.  The Fock-sector superscript has been
dropped, the kinetic and potential energy terms combined into a
single Hamiltonian matrix, and the eigenvalue is $\xi=M^2/\mu^2$.  
The standard approach to such a problem
is to factorize $B$ and convert the problem into
an ordinary eigenvalue problem.  The usual factorization, into a product
of a lower triangular matrix and its transpose, can fail in practice
due to round-off errors in what is an implicit orthogonalization of the
basis.  A reliable factorization is a singular-value decomposition (SVD)
in the form $B=UDU^T$.  The columns of the matrix $U$, and the rows of its
transpose $U^T=U^{-1}$, are the eigenvectors of $B$.  The matrix $D$ is 
diagonal, with the corresponding eigenvalues of $B$ as entries.  
In exact arithmetic, the eigenvalues
must be positive because $B$, as an overlap matrix between basis
functions, is a symmetric positive-definite matrix.  In practice, 
round-off errors can produce small negative eigenvalues; however,
unlike the ordinary factorization, this does not cause the SVD 
factorization to fail.  Instead one can proceed with care.

To incorporate the presence of spurious negative singular values
for $B$, we write $D$ as $|D|^{1/2}S|D|^{1/2}$, with $|D|$ the
absolute value of $D$ and $S$ a diagonal matrix of the signs
of the entries in $D$.  This allows us to define a new vector
$\vec c^{\,\prime}=S|D|^{1/2}U^T\vec c$ and a new matrix
$H'=|D|^{-1/2}U^T H U|D|^{-1/2}S$, such that the eigenvalue
problem becomes an ordinary one: 
$H'\vec c^{\,\prime}=\xi \vec c^{\,\prime}$.  The
remaining complication is that $H'$ is not symmetric;
it is however self-adjoint with respect to the indefinite
metric defined by $S$: $H^{'\dagger}\equiv S H^{\prime T}S^{-1}=H'$.
Of course, for cases when $S$ is strictly positive, we have
$S=I$, and $H'$ is symmetric.  When not, we can use standard
diagonalization for asymmetric matrices, which was found to
work quite well.

\subsection{Convergence}

\begin{figure}[ht]
\vspace{0.2in}
\centerline{\includegraphics[width=15cm]{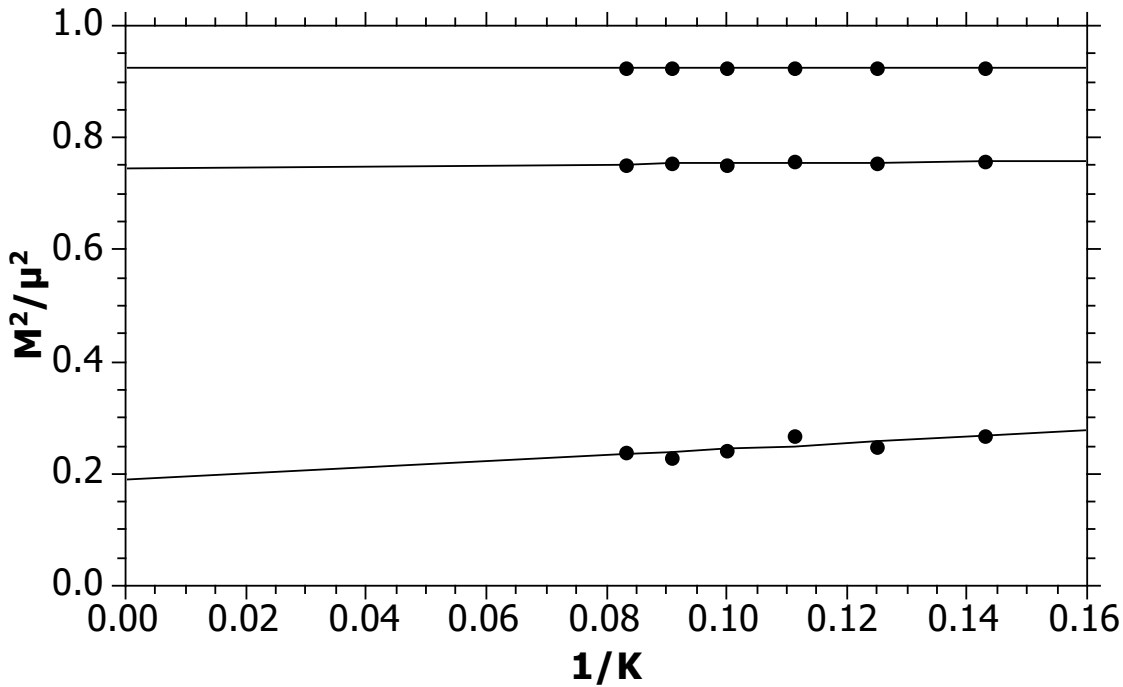}}
\caption{\label{fig:M2extrap}
Lowest mass for the odd case as function of the reciprocal
of the highest order $K$ of the polynomial included in the
five-body Fock sector, for $g=0.5$, 1.0, and 2.0.  In the 
three-body sector, the highest order was 15, and the Fock space 
was truncated at five constituents.  The lines are linear 
fits extrapolated to infinite order.
}
\end{figure}

\begin{figure}[ht]
\vspace{0.2in}
\centerline{\includegraphics[width=15cm]{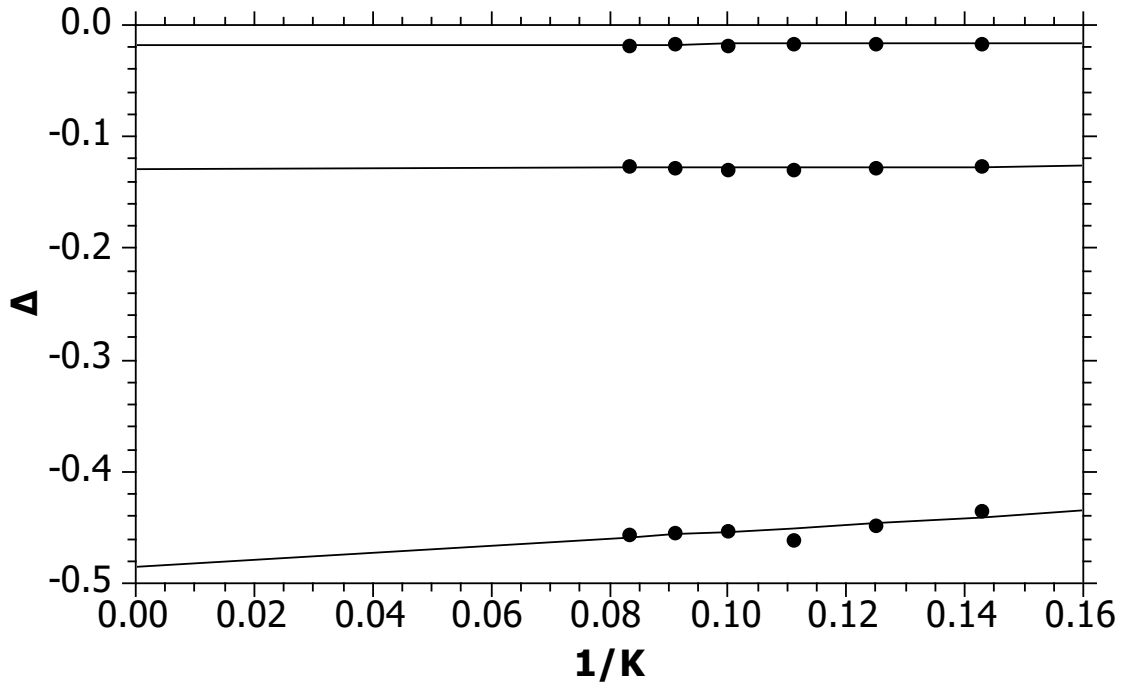}}
\caption{\label{fig:shiftextrap}
Same as Fig.~\ref{fig:M2extrap}, but for the mass renormalization
shift $\Delta$ and $g$ values of 0.5, 1.0, and 1.5.
}
\end{figure}

The convergence with respect to the highest order $K$ of
polynomials in the basis was quite rapid.  Sample extrapolations
are illustrated in Figs.~\ref{fig:M2extrap} and \ref{fig:shiftextrap}.
To get a final number for the mass eigenvalues, for a given
Fock-space truncation, we performed a sequence of such extrapolations, 
varying the highest order in each Fock sector.  With the highest
order in the lower sectors fixed, the highest order in the top
sector was varied and the results extrapolated.  This was repeated for
a range of highest orders in the next lower sector, with
these results each extrapolated.  This layer of extrapolations
was again repeated, until a range of orders had been considered
in every Fock sector. The ranges considered were 10-15 in the
three-body sector, 7-12 in the five body sector, 4-10 in the
seven-body sector, 10-14 in the four-body sector, 6-9 in the
six-body sector, and 4-9 in the eight-body sector, with the
two-body sector fixed at a highest order of 20.
For the shift calculation, the Fock-space truncation was
at five constituents and extrapolation was done only in
that sector, using a range of 7-12, with the highest order
in the three-body sector being 15.

Each extrapolation included an error estimate
in the infinite-order intercept, and for all
but the initial, top-level extrapolation, subsequent
extrapolation was done with contributions weighted by
their errors.  The last extrapolation then yielded an
overall error estimate for the final extrapolated value,
and this was used for error bars in the plots of mass
values.  As the coupling approached the critical
value, the errors grew, as would be expected,
because higher Fock states become more important
for the calculation, leading to a greater dependence
on the basis size in higher sectors.


\end{document}